\pdfoutput=1
\documentclass[sigconf,natbib=true]{acmart}

\usepackage{multirow}
\usepackage{subcaption}
\usepackage{enumitem}
\usepackage{cleveref}
\crefname{figure}{Figure}{Figures}
\Crefname{figure}{Figure}{Figures}
\crefname{table}{Table}{Tables}
\Crefname{table}{Table}{Tables}
\crefname{equation}{Equation}{Equations}
\Crefname{equation}{Equation}{Equations}
\crefname{section}{Section}{Sections}
\Crefname{section}{Section}{Sections}
\usepackage{colortbl}
\usepackage{framed}  
\usepackage{fontawesome5}
\usepackage{xspace}

\newcommand{\system}{SMD\xspace}

\definecolor{tblBand}{HTML}{F2F4F7}
\definecolor{tblRule}{HTML}{2F2F2F}

\definecolor{takeBg}{HTML}{F5F7FA}
\definecolor{takeRule}{HTML}{2E7D32}
\newenvironment{tcbtakeaway}[1]{%
  \par\vspace{2pt}%
  \MakeFramed{\advance\hsize-\width \FrameRestore}%
  \noindent{\small\textbf{\color{takeRule}#1.}\hspace{0.4em}}%
  \small\ignorespaces}{%
  \endMakeFramed\par\vspace{2pt}}

\AtBeginDocument{%
  }

\hypersetup{colorlinks=true, urlcolor=blue, linkcolor=black, citecolor=black}

\setcopyright{none}
\renewcommand\footnotetextcopyrightpermission[1]{}
\copyrightyear{2026}
\acmYear{2026}
\acmConference[CIKM '26]{The 35th ACM International Conference on Information and Knowledge Management}{November 7--11, 2026}{Rome, Italy}

\begin{document}

\title[Sequential Modality Dropout]{Sequential Modality Dropout for Robust Multi-Modal Sequential Recommendation}
\titlenote{Accepted at the 35th ACM International Conference on Information and Knowledge Management (CIKM '26), November 7--11, 2026, Rome, Italy. This is the authors' preprint version.}

\author{Guanqun Yang}
\email{guanqun.yang@outlook.com}
\affiliation{%
  \institution{Stevens Institute of Technology}
  \city{Hoboken}
  \state{NJ}
  \country{USA}}

\author{Wenlong Zhang}
\email{wzhang71@stevens.edu}
\affiliation{%
  \institution{Stevens Institute of Technology}
  \city{Hoboken}
  \state{NJ}
  \country{USA}}

\renewcommand{\shortauthors}{Yang and Zhang}

\begin{abstract}
Multi-modal sequential recommenders assume every item carries every modality, but real product catalogs often miss images or text, and a model trained on complete data loses much of its recommendation accuracy when a modality is unavailable at serving time.
We propose \textbf{Sequential Modality Dropout} (\system): during training, each modality stream (image and text) is independently erased with probability $p$ for an entire user interaction history, so the model learns to predict the next item without relying on any single modality.
We measure robustness by retention, the fraction of a model's full-modality accuracy (HR@10) that survives when a modality is removed at test time.
Across four backbones (MM-SASRec, IISAN, MISSRec, and fMRLRec) on four Amazon domains, \system raises text retention by 1.0 to 3.2$\times$ at essentially no cost to full-modality accuracy; under an extreme 95\% per-item missing rate, it retains 61\% of HR@10 versus 22\% without (a 2.8$\times$ improvement).
An optional cross-modal reconstruction loss further lifts retention from 90\% to 98\% on a simple additive backbone under severe text missingness.
\system is a four-line, architecture-agnostic change that makes multi-modal sequential recommenders robust to the missing modalities they actually encounter in deployment.

\smallskip\noindent\textbf{\faGithub\ Code:} \url{https://github.com/guanqun-yang/SMD}
\end{abstract}



\maketitle

\section{Introduction}
\label{sec:intro}

Many recent sequential recommenders incorporate frozen content features, such as product images encoded by CLIP or ViT and text encoded by BERT or Llama, alongside or instead of the learned ID embeddings of SASRec~\cite{Kang2018SelfAttentiveSequentialRecommendation}, to capture preferences that IDs alone miss~\cite{Wang2023MISSRecPretrainingTransferring,Fu2024IISANEfficientlyAdapting,Wang2024TrainOnceDeploy,Fu2025MultimodalFusionSparse,Hong2025MTSTRecMultimodalTimeAligned,Hu2024BiVRecBidirectionalViewbased,Fu2025CROSSANEfficientEffective}.
These methods are trained and evaluated on benchmarks in which every item carries every modality, but real product catalogs routinely violate this assumption~\cite{Fu2026BenchmarkingMultimodalLarge}: 34.3\% of Toys \& Games and 48.3\% of Beauty \& Personal Care items have no text description,\footnote{\url{https://amazon-reviews-2023.github.io/} (Amazon Reviews 2023, 10-core filtered; our measurement).} and 26\% to 41\% of items have no image on the four MISSRec Amazon domains.\footnote{\url{https://github.com/gimpong/MM23-MISSRec} (Amazon splits; our measurement).}
Removing text from a vanilla multi-modal SASRec at test time drops HR@10 to 40 to 73\% of its full-modality value (\cref{tab:fullremoval}), and under a 95\% per-item missing rate retention collapses to 22\% (\cref{fig:retention}).

\begin{figure*}[t]
  \centering
  \includegraphics[width=\textwidth]{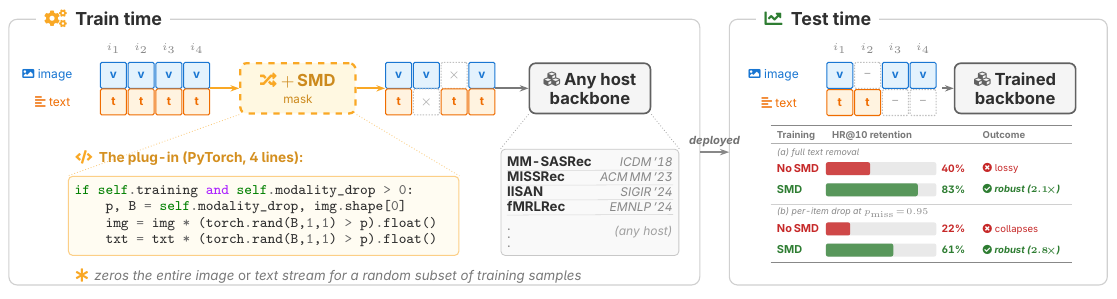}
  \caption{\textbf{Sequential Modality Dropout (\system).} A four-line per-sample Bernoulli modality mask injected at the fusion point of any multi-modal sequential backbone (MM-SASRec~\cite{Kang2018SelfAttentiveSequentialRecommendation}, MISSRec~\cite{Wang2023MISSRecPretrainingTransferring}, IISAN~\cite{Fu2024IISANEfficientlyAdapting}, fMRLRec~\cite{Wang2024TrainOnceDeploy}). At test time, the same trained backbone retains 61\% of HR@10 at $p_{\mathrm{miss}}\!=\!0.95$ versus 22\% for the unmodified model.}
  \label{fig:headline}
\end{figure*}

Missing modalities, however, have been studied almost entirely outside the sequential setting.
Within recommendation, the problem is treated as a collaborative-filtering (CF) one: FeatProp~\cite{Malitesta2024DealingMissingModalities} propagates features along an item-to-item co-interaction graph, MMGACL~\cite{Zhao2025GraphAttentionContrastive} completes features by diffusion with bimodal attention, SiBraR~\cite{Ganhor2024MultimodalSingleBranchEmbedding} trains a single-branch encoder on random modality subsets, and LRMM~\cite{Wang2018LRMMLearningRecommend} applies modality dropout to review-based rating prediction, none of which model the temporal item ordering that sequential recommenders exploit.
Outside recommendation, Bernoulli modality masking is a mature training-time tool, used for audiovisual gesture recognition~\cite{Neverova2015ModDropAdaptiveMultimodal}, talking-face synthesis~\cite{Abdelaziz2020ModalityDropoutImproved}, medical-image segmentation~\cite{Liu2022ModDropDynamicFilter}, and masked cross-modal projection~\cite{Nezakati2024MMPRobustMultiModal}, but always on a single sample, where no user sequence exists.
Neither line covers multi-modal \emph{sequential} recommendation, a gap that a recent survey of 354 missing-modality papers~\cite{Wu2026DeepMultimodalLearning} names explicitly as an under-explored temporal setting.
We close it by bringing modality masking into sequential recommendation with a design built for the temporal structure of user histories.

We propose \textbf{Sequential Modality Dropout} (\system; \cref{fig:headline}), a per-sample Bernoulli modality mask injected at the fusion point of any multi-modal sequential recommender.
For each training sample, each modality stream (image and text) is independently zeroed with probability $p$, and the same mask applies to every item in the user's chronological sequence.
This design preserves the temporal signal the sequential Transformer relies on, and matches real catalog missingness, which tends to cluster within a session or a product category (e.g., image-CDN outages) rather than varying independently across items.
\system is also a \emph{black-box plug-in}: the same module integrates into MM-SASRec, IISAN~\cite{Fu2024IISANEfficientlyAdapting}, MISSRec~\cite{Wang2023MISSRecPretrainingTransferring}, and fMRLRec~\cite{Wang2024TrainOnceDeploy} without per-architecture tuning and leaves the host optimizer and hyperparameters untouched.

Our contributions answer three research questions, examined in turn in \cref{sec:exp}:
\begin{itemize}[leftmargin=1.2em,itemsep=1pt,topsep=2pt]
\item \textbf{A universal plug-in across backbones (RQ1).} On Amazon Scientific across the four backbones, \system lifts HR@10 text retention from 18--94\% to 56--99\% (\cref{tab:plugin}); on MM-SASRec across four Amazon domains, it lifts text retention from 40--73\% to 79--97\% (\cref{tab:fullremoval}). Across all 11 of the 16 possible (backbone, dataset) combinations, the robustness gain is 1.0 to 3.2$\times$ at essentially no cost to full-modality accuracy.
\item \textbf{Robustness that scales to extreme missingness (RQ2).} On a 0 to 95\% per-item missing-rate sweep, MM-SASRec with \system retains 61\% of HR@10 at $p_{\mathrm{miss}}\!=\!0.95$ versus 22\% for the unmodified model (\cref{fig:retention}), and per-user paired tests on 121k users confirm the gains are statistically significant (unlikely to arise by chance).
\item \textbf{An opt-in cross-modal reconstruction loss (RQ3).} For simple additive backbones under severe text missingness, an auxiliary loss that trains the two modality projections to predict each other lifts text retention from 90\% to 98\% on Beauty \& Personal Care (48\% text-missing; \cref{tab:recon}).
\end{itemize}


\section{Sequential Modality Dropout}
\label{sec:method}

\system has three parts, described in turn: the multi-modal backbone it plugs into (\cref{sec:method-backbone}), the per-sample modality-masking mechanism at its core (\cref{sec:method-smd}), and an optional cross-modal reconstruction loss for severe missingness (\cref{sec:recon}).

\subsection{Multi-Modal Sequential Backbone}
\label{sec:method-backbone}

We work with the SASRec backbone~\cite{Kang2018SelfAttentiveSequentialRecommendation} extended with the additive multi-modal fusion of \cref{eq:fusion} (we refer to this construction as MM-SASRec, following the multi-modal extension used in MISSRec~\cite{Wang2023MISSRecPretrainingTransferring} and IISAN~\cite{Fu2024IISANEfficientlyAdapting}).
Each item $i$ has a learned ID embedding $\mathbf{e}_i^{\mathrm{ID}} \in \mathbb{R}^d$ and two frozen content vectors: an image embedding $\mathbf{v}_i$ from a CLIP visual encoder and a text embedding $\mathbf{t}_i$ from a language encoder.
Two linear projections $f_v$ and $f_t$ map the content vectors into the model's $d$-dimensional hidden space, and the per-item representation is the additive fusion
\begin{equation}
\label{eq:fusion}
\mathbf{e}_i = \mathbf{e}_i^{\mathrm{ID}}
   + a_i^{v}\, f_v(\mathbf{v}_i)
   + a_i^{t}\, f_t(\mathbf{t}_i),
\end{equation}
where $a_i^{v}, a_i^{t} \in \{0,1\}$ are catalog availability indicators (a missing modality contributes zero).
The sequence $\mathbf{e}_{i_1}, \ldots, \mathbf{e}_{i_{n-1}}$ is fed to a causal Transformer block stack and a linear output head scores all items in the catalog.
The training objective is the standard binary cross-entropy loss with one positive and one sampled negative per position.

\subsection{Modality Masking}
\label{sec:method-smd}

\paragraph{Mechanism.}
\system inserts a per-sample Bernoulli mask on the modality streams immediately before fusion (\cref{fig:headline}).
During training, we draw an independent Bernoulli mask for each sample $b$ in the minibatch and each modality $m \in \{v,t\}$,
\begin{equation}
\label{eq:smd}
m_b^{(m)} \sim \mathrm{Bernoulli}(1-p),
\quad
\tilde{f}_m(\mathbf{x}_i^{(m)}) = m_b^{(m)} \cdot
   f_m(\mathbf{x}_i^{(m)}),
\end{equation}
and substitute $\tilde f_m$ for $f_m$ in \cref{eq:fusion}.
The mask is per-sample, not per-item: all items in user $b$'s sequence share the same modality mask.
This matches how modalities go missing in real catalogs, where an entire product category or data source tends to lack a modality at once, rather than an artificial pattern where missing items are scattered evenly through the sequence (for example, a category whose text descriptions are absent in bulk, or an image feed unavailable for a whole batch of items).
At test time the mask is not applied, except for the deterministic masks used in our robustness evaluation (\cref{sec:exp-setup}).

\paragraph{Relation to Standard Dropout.}
Unlike standard dropout, which at test time rescales activations by $1/(1-p)$ to preserve their expected magnitude as a regularizer~\cite{SrivastavaDropoutSimpleWay}, \system zeroes a whole modality with no rescaling, following the established modality-dropout convention~\cite{Neverova2015ModDropAdaptiveMultimodal,Abdelaziz2020ModalityDropoutImproved}.
Its goal is invariance to a genuinely missing modality rather than variance reduction, so the full-modality input seen at test is simply the $p\!=\!0$ case the model already encountered during training, and no compensating scale is required.

\paragraph{Why Per-Sample.}
A natural alternative is a \emph{per-item} mask that decides separately for each item whether to drop a modality, so some items in a sequence keep both modalities while others lose one.
We use the per-sample mask instead because real missingness is whole-modality and structured by category or source rather than independent across items: a modality tends to go absent in a block, the same pattern that produces the cold-start problem~\cite{Wang2018LRMMLearningRecommend,Ganhor2024MultimodalSingleBranchEmbedding}, and in our data the text-missing rate varies sharply across categories (34.3\% to 48.3\%).
A per-item mask would instead train the model for an independent, missing-at-random pattern that is rarely seen in deployment~\cite{Wang2018LRMMLearningRecommend}.

\paragraph{Implementation.}
The entire mechanism is the four-line modification at the fusion point of the host model shown in \cref{fig:headline}.
We apply the same template at the corresponding line in each of MM-SASRec, IISAN, MISSRec, and fMRLRec; the only per-architecture decision is which tensor in the fusion path carries the unpooled modality embeddings.

\subsection{Cross-Modal Reconstruction}
\label{sec:recon}

For backbones whose fusion layer offers no native fallback path (e.g., the simple additive fusion in \cref{eq:fusion}), we additionally consider a small cross-modal reconstruction loss in the spirit of MMP's masked modality projection~\cite{Nezakati2024MMPRobustMultiModal}
\begin{equation}
\label{eq:recon}
\mathcal{L}_{\mathrm{rec}} =
   \tfrac{1}{2}\!\left(
     \|g_{t\to v}(f_t(\mathbf{t}_i)) - f_v(\mathbf{v}_i)\|^2
   + \|g_{v\to t}(f_v(\mathbf{v}_i)) - f_t(\mathbf{t}_i)\|^2
   \right),
\end{equation}
where $g_{t\to v}$ and $g_{v\to t}$ are two-layer projections trained jointly with the recommender.
The total objective is $\mathcal{L} = \mathcal{L}_{\mathrm{BCE}} + \lambda\, \mathcal{L}_{\mathrm{rec}}$, where we set $\lambda = 0.01$ from preliminary runs; a larger $\lambda = 0.1$ reduced full-modality accuracy.
The intuition is that, when modality $m$ is masked at test time, the model can implicitly recover an estimate of it through the surviving modality.
We show in \cref{sec:exp-rq3} that this auxiliary loss is worth its accuracy cost only on simple additive backbones in extreme-missingness regimes.

\section{Experiments}
\label{sec:exp}

We evaluate \system on the four-domain MISSRec benchmark, organizing the study around three questions:
\begin{description}[leftmargin=2.4em,itemsep=1pt,topsep=2pt,labelindent=0pt]
\item[RQ1] Does \system work across architectures?
\item[RQ2] How does \system scale with the missing rate?
\item[RQ3] Can a cross-modal reconstruction loss boost \system further?
\end{description}

\subsection{Setup}
\label{sec:exp-setup}

\paragraph{Datasets.}
We used four Amazon domains following the MISSRec benchmark~\cite{Wang2023MISSRecPretrainingTransferring}: Scientific, Instruments, Arts, and Office, spanning 4,385 to 25,986 items, with 26\% to 41\% image-missing rates measured on our re-downloaded catalog.
All splits were chronological, with a maximum sequence length of 10, and full-catalog ranking (each held-out item is scored against every catalog item, not a sampled subset).
The reconstruction-loss study also used Beauty \& Personal Care from the Amazon Reviews 2023 release, 10-core filtered, with 48.3\% of items missing text in our measurement.

\paragraph{Backbones.}
We evaluated \system on four backbones spanning the design space: \textbf{MM-SASRec} (additive fusion of \cref{eq:fusion} on a 1-block hidden-64 SASRec~\cite{Kang2018SelfAttentiveSequentialRecommendation}), \textbf{IISAN}~\cite{Fu2024IISANEfficientlyAdapting} (frozen ViT and BERT encoders adapted by small trainable side-networks, i.e., parameter-efficient fine-tuning), \textbf{MISSRec}~\cite{Wang2023MISSRecPretrainingTransferring} (a Transformer with dynamic modality fusion, fine-tuned from a 100-epoch cross-domain checkpoint), and \textbf{fMRLRec}~\cite{Wang2024TrainOnceDeploy} (an efficient linear-recurrent, state-space sequence model whose Matryoshka embeddings can be truncated to smaller sizes).
For each backbone, we trained two checkpoints (one without \system and one with \system at $p\!=\!0.3$) using identical hyperparameters, optimizers, and seeds, sharing frozen MISSRec CLIP ViT-B/32 features (512-dim) across backbones; \system is the \emph{only} difference between the two.

\paragraph{Evaluation Protocols.}
A single test-time missingness pattern can over- or under-state robustness, so we use two complementary protocols:
\begin{itemize}[leftmargin=1.2em,itemsep=1pt,topsep=2pt]
\item \textit{Protocol~1} (full-modality removal) zeroes one or both modalities for every item at test time, yielding the conditions \texttt{full}, \texttt{no\_text}, \texttt{no\_image}, and \texttt{no\_modal}; it matches a categorical or systemic outage in which an entire modality is unavailable~\cite{Ganhor2024MultimodalSingleBranchEmbedding}.
\item \textit{Protocol~2} (per-item drop) zeroes each modality of each item independently with probability $p_{\mathrm{miss}} \in [0, 0.95]$, averaged over 5 random seeds, matching item-level missingness in real catalogs~\cite{Malitesta2024DealingMissingModalities,Fu2026BenchmarkingMultimodalLarge}.
\end{itemize}
Together they cover the worst case (every item missing the same modality) and the realistic average case (items missing modalities at random).
For each checkpoint we report HR@10 ($\times 100$) and the \emph{retention rate} $R$; comparing a model trained with versus without \system gives the \emph{robustness gain} $G$:
\begin{equation}
\label{eq:rg}
R(\theta) = \frac{\mathrm{HR@10}(\theta;\,\text{missing})}{\mathrm{HR@10}(\theta;\,\text{full})},
\quad
G = \frac{R(\theta_{\mathrm{SMD}})}{R(\theta_{\mathrm{No\,SMD}})},
\end{equation}
where $R(\theta)$ is defined for a single trained checkpoint $\theta$ and $G$ compares the two checkpoints trained with matched hyperparameters and seeds.

\subsection{RQ1: \system Works Across Architectures}
\label{sec:exp-rq1}

\begin{table*}[t]
\centering
\footnotesize
\caption{Results (HR@10 $\times 100$). The retention rate is $R = \text{HR@10}(\text{missing})/\text{HR@10}(\text{full})$ for the removed modality, and the robustness gain is $G = R_{\text{SMD}}/R_{\text{No SMD}}$; dropout is $p\!=\!0.3$ unless noted. (a) \system as a plug-in across four backbones; (b) MM-SASRec across four domains; (c) the optional cross-modal reconstruction loss.}
\label{tab:rq1}
\begin{subtable}[t]{0.31\textwidth}
\centering
\setlength{\tabcolsep}{3pt}
\renewcommand{\arraystretch}{1.05}
\caption{Plug-in across four backbones (Scientific, text removed); best text retention per backbone in bold.}
\label{tab:plugin}
\begin{tabular}{@{}llrrrc@{}}
\toprule
\textbf{Backbone} & \textbf{Train.} & \textbf{Full} & \textbf{No\,txt} & $\boldsymbol{R}_{\text{text}}$ & $\boldsymbol{G}$ \\
\midrule
\multirow{2}{*}{IISAN~\cite{Fu2024IISANEfficientlyAdapting}}
 & --   & 6.43  & 1.13 & 18\% & --          \\
 & SMD  & 6.15  & 3.44 & \textbf{56\%} & $3.2\times$ \\
\cmidrule(lr){1-6}
\multirow{2}{*}{MISSRec~\cite{Wang2023MISSRecPretrainingTransferring}}
 & --   & 13.53 & 5.80 & 43\% & --          \\
 & SMD  & 13.27 & 9.31 & \textbf{70\%} & $1.6\times$ \\
\cmidrule(lr){1-6}
\multirow{2}{*}{fMRLRec~\cite{Wang2024TrainOnceDeploy}}
 & --   & 4.83  & 4.54 & 94\% & --          \\
 & SMD  & 4.79  & 4.75 & \textbf{99\%} & $1.1\times$ \\
\cmidrule(lr){1-6}
\multirow{2}{*}{MM-SASRec~\cite{Kang2018SelfAttentiveSequentialRecommendation}}
 & --   & 6.18  & 2.45 & 40\% & --          \\
 & SMD  & 6.31  & 5.22 & \textbf{83\%} & $2.1\times$ \\
\bottomrule
\end{tabular}

\end{subtable}\hfill
\begin{subtable}[t]{0.37\textwidth}
\centering
\setlength{\tabcolsep}{3pt}
\renewcommand{\arraystretch}{1.05}
\caption{MM-SASRec across four domains (text or image removed); the better No-SMD/SMD value per cell in bold.}
\label{tab:fullremoval}
\begin{tabular}{@{}llrrrrr@{}}
\toprule
\textbf{Dataset} & \textbf{Train.} & \textbf{Full} & \textbf{No\,txt} & \textbf{No\,img} & $\boldsymbol{R}_{\text{txt}}$ & $\boldsymbol{R}_{\text{img}}$ \\
\midrule
\multirow{2}{*}{Scientific}
 & --  & 6.18 & 2.45 & 5.39 & 40\% & 87\% \\
 & SMD & \textbf{6.31} & \textbf{5.22} & \textbf{5.93} & \textbf{83\%} & \textbf{94\%} \\
\cmidrule(lr){1-7}
\multirow{2}{*}{Instruments}
 & --  & 7.14 & 5.18 & 6.34 & 73\% & 89\% \\
 & SMD & 7.13 & \textbf{6.95} & \textbf{6.85} & \textbf{97\%} & \textbf{96\%} \\
\cmidrule(lr){1-7}
\multirow{2}{*}{Arts}
 & --  & 4.75 & 2.32 & 3.68 & 49\% & 77\% \\
 & SMD & \textbf{5.06} & \textbf{4.01} & \textbf{4.67} & \textbf{79\%} & \textbf{92\%} \\
\cmidrule(lr){1-7}
\multirow{2}{*}{Office}
 & --  & 5.30 & 2.63 & 3.57 & 50\% & 67\% \\
 & SMD & \textbf{5.62} & \textbf{4.62} & \textbf{4.89} & \textbf{82\%} & \textbf{87\%} \\
\bottomrule
\end{tabular}

\end{subtable}\hfill
\begin{subtable}[t]{0.30\textwidth}
\centering
\setlength{\tabcolsep}{3pt}
\renewcommand{\arraystretch}{1.05}
\caption{Reconstruction loss (MISSRec on Scientific, MM-SASRec on Beauty); best No-text HR@10 and retention in bold.}
\label{tab:recon}
\begin{tabular}{@{}llrrr@{}}
\toprule
\textbf{Backbone} & \textbf{Training} & \textbf{Full} & \textbf{No\,txt} & $\boldsymbol{R}_{\text{text}}$ \\
\midrule
\multirow{3}{*}{MISSRec}
 & --            & 13.53 & 5.80 & 43\% \\
 & SMD           & 13.27 & 9.31 & 70\% \\
 & SMD\,$+$\,recon & 12.96 & \textbf{9.49} & \textbf{73\%} \\
\cmidrule(lr){1-5}
\multirow{4}{*}{MM-SASRec}
 & --                  & 0.72 & 0.31 & 43\% \\
 & SMD ($p\!=\!0.3$)   & 0.63 & 0.57 & 90\% \\
 & SMD ($p\!=\!0.5$)   & 0.56 & 0.48 & 86\% \\
 & SMD\,$+$\,recon       & 0.52 & \textbf{0.51} & \textbf{98\%} \\
\bottomrule
\end{tabular}

\end{subtable}
\end{table*}

\Cref{tab:plugin} reports HR@10 under text removal for the four backbones on Amazon Scientific, where text retention without \system varies most across backbones, from 18\% (IISAN) to 94\% (fMRLRec).
\system improves text retention in every backbone, with gain $G \in [1.1\times,\, 3.2\times]$: the largest lifts are on IISAN (18\% to 56\%, 3.2$\times$) and MM-SASRec (40\% to 83\%, 2.1$\times$), while fMRLRec, already 94\% retained because its concat-plus-projection fusion learns near-redundant modality embeddings, rises only to 99\%.

\Cref{tab:fullremoval} reports MM-SASRec when text and when images are removed across four domains: \system lifts text retention from 40--73\% to 79--97\% and image retention from 67--89\% to 87--96\%, while matching or exceeding peak HR@10 on every dataset.
Text removal is consistently the harder condition because the text encoder carries product-name information that the image cannot recover.

Beyond these two slices, we ran all 16 (backbone, dataset) combinations and summarize them here, omitting the full table only to respect the page limit: 11 of the 16 improve, with robustness gain 1.0 to 3.2$\times$ and a mean peak-HR@10 change of +0.8\% (range -4.4\% to +6.5\%).
The two largest accuracy losses, IISAN/Scientific (-4.4\%) and IISAN/Instruments (-4.1\%), occur exactly where \system delivers its largest retention gains (3.2$\times$ and 1.4$\times$).

\begin{tcbtakeaway}{RQ1 Takeaway}
\system helps on every recommender we tested, regardless of how it fuses image and text; the biggest robustness gains coincide with the biggest full-modality accuracy drops, yet even those drops stay small (at most 4.4\%).
\end{tcbtakeaway}

\subsection{RQ2: \system Scales to Extreme Missingness}
\label{sec:exp-rq2}

\begin{figure}[t]
  \centering
  \includegraphics[width=\columnwidth]{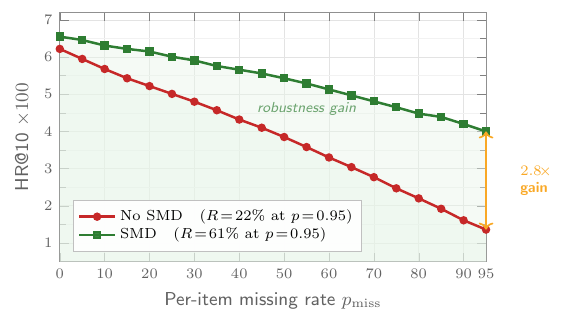}
  \caption{Per-item missing-rate sweep on Scientific (MM-SASRec, mean of 5 seeds). At $p_{\mathrm{miss}}\!=\!0.95$, \system retains 61\% of HR@10 versus 22\% without.}
  \label{fig:retention}
\end{figure}

\Cref{fig:retention} sweeps $p_{\mathrm{miss}}$ from 0\% to 95\% in 5\% increments on Scientific.
The two curves diverge \emph{monotonically} from $p_{\mathrm{miss}}\!=\!0$: the \system curve is approximately flat (slope $\approx -0.027$ HR@10 per unit $p_{\mathrm{miss}}$), while the No-\system curve falls steeply (slope $\approx -0.051$); at $p_{\mathrm{miss}}\!=\!0.95$, \system retains 61\% versus 22\% for the same model without \system, a 2.8$\times$ improvement.
The same pattern holds on Arts and Office: at 70\% per-item drop \system retains 71--78\% versus 46--53\% without \system; we plot only the Scientific sweep (\cref{fig:retention}) and omit the per-domain tables to respect the page limit.

To rule out user-level noise, we pair each user's HR@10 and NDCG@10 between the No-\system and \system checkpoints across three datasets (121k users).
Because HR@10 is binary per user, we test it with McNemar's test, which weighs how many users \system flips from miss to hit against the reverse; NDCG@10 is a continuous score (higher when the held-out item is ranked nearer the top), so we use the Wilcoxon signed-rank test, which ranks the per-user differences and assumes no particular distribution.
Of the resulting 24 tests (12 (condition, dataset) pairs $\times$ 2 tests), 22 are significant at $\alpha = 0.05$, meaning a gap this large is very unlikely if the two models were equivalent, with $p$-values from $4.3\times 10^{-3}$ to $1.0\times 10^{-233}$; the two exceptions are the full-modality McNemar tests on Scientific and Instruments, where \system flips similar numbers of users each way.
We ran all 24 tests and report these aggregates in place of the full table, which the page limit does not permit us to include.

\begin{tcbtakeaway}{RQ2 Takeaway}
Robustness scales with missingness: the gap between \system and the unmodified model widens monotonically and remains statistically significant across datasets and conditions.
\end{tcbtakeaway}

\subsection{RQ3: A Cross-Modal Reconstruction Loss for Severe Missingness}
\label{sec:exp-rq3}

The auxiliary loss $\mathcal{L}_{\mathrm{rec}}$ (\cref{eq:recon}) trains the two modality projections to predict each other, so the model can implicitly recover a missing modality at test time.
Its value depends sharply on the backbone (\cref{tab:recon}).
On a strong dynamic-fusion backbone (MISSRec on Scientific) it adds only 3 points of text retention (70\% to 73\%) while costing 2.3\% of full-modality HR@10, a net loss across most metrics MISSRec's own results~\cite{Wang2023MISSRecPretrainingTransferring} report.
On a simple additive backbone under severe missingness (MM-SASRec on Beauty \& Personal Care, 48\% text-missing), it instead lifts retention from 90\% to 98\%, because the model has no dynamic fusion to fall back on.

\begin{tcbtakeaway}{RQ3 Takeaway}
The reconstruction loss helps only when fusion is simple and missingness is severe; \system alone is the default, with the loss an opt-in enhancement.
\end{tcbtakeaway}

\section{Conclusion}
\label{sec:conclusion}

\system is a four-line, per-sample training-time mask at the fusion point of any multi-modal sequential recommender, addressing the accuracy loss that missing modalities cause at serving time.
Across four backbones and four Amazon domains, it lifts HR@10 text retention by 1.0 to 3.2$\times$ over 11 of 16 combinations at +0.8\% mean peak HR@10, and under an extreme 95\% per-item missing rate it retains 61\% of HR@10 versus 22\% without.
Per-user tests on 121k users confirm these gains are statistically significant.
An optional cross-modal reconstruction loss lifts retention from 90\% to 98\% on simple additive backbones under severe missingness.
Because \system is a self-contained change at the fusion step, it can be combined with complementary techniques such as feature propagation~\cite{Malitesta2024DealingMissingModalities} and cross-modal reconstruction~\cite{Nezakati2024MMPRobustMultiModal}.
Two directions remain open: extending beyond image and text to audio, video, or structured attributes, and handling modality \emph{corruption} rather than absence.

\newpage
\section*{GenAI Usage Disclosure}

The authors used Claude Code and the Gemini CLI only for technical execution, such as generating code for diagrams and plots, and for polishing the manuscript's readability.
The foundational research, including algorithm design, methodology, and the initial draft, was entirely the authors' own intellectual work. No AI tools were used to formulate research ideas, create primary data, or execute evaluations.

\bibliographystyle{ACM-Reference-Format}
\bibliography{zotero}

\appendix

\section{Dataset Statistics}
\label{app:datasets}

\Cref{tab:datasets} reports the per-domain catalog size, total number of interactions, and image-missing rate of our re-downloaded MISSRec catalog used in the main experiments; these numbers differ from the coverage figures in MISSRec's original Table 1 (Scientific 26.75\%, Pantry 93.65\%, Instruments 63.12\%, Arts 44.90\%, Office 63.99\%) because we re-downloaded images directly from the Amazon Reviews 2023 release rather than reusing the MISSRec-provided archives.

\begin{table}[tb]
\centering
\caption{MISSRec benchmark datasets. ``Image-miss.'' is the fraction of catalog items whose image is unavailable in our re-download from the Amazon Reviews 2023 release; the Office rate was not recorded at download time and is marked ``n/a''.}
\label{tab:datasets}
\small
\setlength{\tabcolsep}{4.5pt}
\renewcommand{\arraystretch}{1.12}
\begin{tabular}{@{}lrrr@{}}
\toprule
\textbf{Dataset} & \textbf{\#\,Items} & \textbf{\#\,Inter.} & \textbf{Image-miss.} \\
\midrule
Scientific  &  4{,}385 &  51\,k & 32.8\% \\
Instruments &  9{,}964 & 134\,k & 25.7\% \\
Arts        & 21{,}019 & 259\,k & 40.9\% \\
Office      & 25{,}986 & 310\,k & \multicolumn{1}{c}{n/a} \\
\bottomrule
\end{tabular}
\end{table}

\section{Full Plug-In Robustness Results}
\label{app:plugin-full}

\Cref{tab:plugin-full} reports the full SMD plug-in results across the four backbones and three Amazon domains referenced in \cref{sec:exp-rq1}.
The condensed Scientific-only view in \cref{tab:plugin} of the main paper is a slice of this table.

\begin{table}[tb]
\centering
\caption{Full SMD plug-in results across four backbones and three Amazon domains. HR@10$\times 100$. Best $R_{\text{text}}$ per pair in bold. Arts runs are omitted for IISAN, MISSRec, and fMRLRec for the same compute-budget reason; the MM-SASRec/Arts cell is the $11$th cell in our main claim and appears in \cref{tab:fullremoval}. Office runs are omitted for MISSRec and fMRLRec because their training cost on the larger Office catalog ($25{,}986$ items, $310$k interactions) is prohibitive at our compute budget; the trend holds on Office for IISAN ($34\%\!\to\!77\%$) and MM-SASRec ($50\%\!\to\!82\%$).}
\label{tab:plugin-full}
\small
\setlength{\tabcolsep}{4.5pt}
\renewcommand{\arraystretch}{1.12}
\begin{tabular}{@{}llcrrrc@{}}
\toprule
\textbf{Backbone} & \textbf{Dataset} & \textbf{Train.} & \textbf{Full} & \textbf{No\,txt} & $\boldsymbol{R}_{\text{text}}$ & $\boldsymbol{G}$ \\
\midrule
\multirow{6}{*}{IISAN~\cite{Fu2024IISANEfficientlyAdapting}}
 & \multirow{2}{*}{Scientific}  & --   & 6.43 & 1.13 & 18\% & --       \\
 &                              & SMD  & 6.15 & 3.44 & \textbf{56\%} & $3.2\times$ \\
\cmidrule(lr){2-7}
 & \multirow{2}{*}{Instruments} & --   & 8.74 & 5.17 & 59\% & --       \\
 &                              & SMD  & 8.38 & 6.87 & \textbf{82\%} & $1.4\times$ \\
\cmidrule(lr){2-7}
 & \multirow{2}{*}{Office}      & --   & 6.46 & 2.19 & 34\% & --       \\
 &                              & SMD  & 6.59 & 5.05 & \textbf{77\%} & $2.3\times$ \\
\midrule
\multirow{4}{*}{MISSRec~\cite{Wang2023MISSRecPretrainingTransferring}}
 & \multirow{2}{*}{Scientific}  & --   & 13.53 & 5.80 & 43\% & --      \\
 &                              & SMD  & 13.27 & 9.31 & \textbf{70\%} & $1.6\times$ \\
\cmidrule(lr){2-7}
 & \multirow{2}{*}{Instruments} & --   & 12.92 & 7.10 & 55\% & --      \\
 &                              & SMD  & 12.85 & 9.71 & \textbf{76\%} & $1.4\times$ \\
\midrule
\multirow{4}{*}{fMRLRec~\cite{Wang2024TrainOnceDeploy}}
 & \multirow{2}{*}{Scientific}  & --   & 4.83 & 4.54 & 94\% & --       \\
 &                              & SMD  & 4.79 & 4.75 & \textbf{99\%} & $1.1\times$ \\
\cmidrule(lr){2-7}
 & \multirow{2}{*}{Instruments} & --   & 5.07 & 4.82 & 95\% & --       \\
 &                              & SMD  & 5.27 & 5.05 & \textbf{96\%} & $1.0\times$ \\
\midrule
\multirow{6}{*}{MM-SASRec~\cite{Kang2018SelfAttentiveSequentialRecommendation}}
 & \multirow{2}{*}{Scientific}  & --   & 6.18 & 2.45 & 40\% & --       \\
 &                              & SMD  & 6.31 & 5.22 & \textbf{83\%} & $2.1\times$ \\
\cmidrule(lr){2-7}
 & \multirow{2}{*}{Instruments} & --   & 7.14 & 5.18 & 73\% & --       \\
 &                              & SMD  & 7.13 & 6.95 & \textbf{97\%} & $1.3\times$ \\
\cmidrule(lr){2-7}
 & \multirow{2}{*}{Office}      & --   & 5.30 & 2.63 & 50\% & --       \\
 &                              & SMD  & 5.62 & 4.62 & \textbf{82\%} & $1.7\times$ \\
\bottomrule
\end{tabular}
\end{table}

\section{Per-Item Missingness Curve}
\label{app:peritem}

\Cref{tab:peritem} reports the full per-item missing-rate sweep on Scientific (MM-SASRec) used to draw \cref{fig:retention} of the main paper, with values at every $5\%$ step.
\Cref{tab:peritem-arts} and \cref{tab:peritem-office} report coarser four-point sweeps on Arts and Office ($p_{\mathrm{miss}}\in\{0.10,\,0.30,\,0.50,\,0.70\}$), which show the same qualitative trend: at $p_{\mathrm{miss}}\!=\!0.70$, SMD retains $71$--$78\%$ of HR@10 versus $46$--$53\%$ without SMD.

\begin{table}[tb]
\centering
\caption{Full per-item missing-rate sweep on Scientific (MM-SASRec, mean of $5$ seeds). HR@10$\times 100$.}
\label{tab:peritem}
\small
\setlength{\tabcolsep}{4.5pt}
\renewcommand{\arraystretch}{1.12}
\begin{tabular}{@{}rrrrr@{}}
\toprule
$\boldsymbol{p_{\text{miss}}}$ & \textbf{No\,SMD\,HR} & \textbf{SMD\,HR} & $\boldsymbol{R}$\textbf{(No\,SMD)} & $\boldsymbol{R}$\textbf{(SMD)} \\
\midrule
 0\% & 6.22 & 6.55 & 100\% & 100\% \\
 5\% & 5.95 & 6.46 &  96\% &  99\% \\
10\% & 5.68 & 6.31 &  91\% &  96\% \\
15\% & 5.43 & 6.22 &  87\% &  95\% \\
20\% & 5.22 & 6.15 &  84\% &  94\% \\
25\% & 5.01 & 6.01 &  81\% &  92\% \\
30\% & 4.80 & 5.91 &  77\% &  90\% \\
35\% & 4.57 & 5.76 &  74\% &  88\% \\
40\% & 4.32 & 5.66 &  69\% &  86\% \\
45\% & 4.10 & 5.56 &  66\% &  85\% \\
50\% & 3.85 & 5.43 &  62\% &  83\% \\
55\% & 3.58 & 5.29 &  58\% &  81\% \\
60\% & 3.30 & 5.13 &  53\% &  78\% \\
65\% & 3.04 & 4.97 &  49\% &  76\% \\
70\% & 2.77 & 4.81 &  44\% &  73\% \\
75\% & 2.47 & 4.65 &  40\% &  71\% \\
80\% & 2.20 & 4.48 &  35\% &  68\% \\
85\% & 1.92 & 4.39 &  31\% &  67\% \\
90\% & 1.61 & 4.20 &  26\% &  64\% \\
95\% & 1.36 & 4.00 &  22\% &  61\% \\
\bottomrule
\end{tabular}
\end{table}

\begin{table}[tb]
\centering
\caption{Per-item missing-rate sweep on Arts (MM-SASRec, mean of $5$ seeds). HR@10$\times 100$.}
\label{tab:peritem-arts}
\small
\setlength{\tabcolsep}{4.5pt}
\renewcommand{\arraystretch}{1.12}
\begin{tabular}{@{}rrrrr@{}}
\toprule
$\boldsymbol{p_{\text{miss}}}$ & \textbf{No\,SMD\,HR} & \textbf{SMD\,HR} & $\boldsymbol{R}$\textbf{(No\,SMD)} & $\boldsymbol{R}$\textbf{(SMD)} \\
\midrule
 0\% & 4.82 & 5.17 & 100\% & 100\% \\
10\% & 4.56 & 5.02 &  95\% &  97\% \\
30\% & 3.99 & 4.78 &  83\% &  92\% \\
50\% & 3.32 & 4.44 &  69\% &  86\% \\
70\% & 2.54 & 4.05 &  53\% &  78\% \\
\bottomrule
\end{tabular}
\end{table}

\begin{table}[tb]
\centering
\caption{Per-item missing-rate sweep on Office (MM-SASRec, mean of $5$ seeds). HR@10$\times 100$.}
\label{tab:peritem-office}
\small
\setlength{\tabcolsep}{4.5pt}
\renewcommand{\arraystretch}{1.12}
\begin{tabular}{@{}rrrrr@{}}
\toprule
$\boldsymbol{p_{\text{miss}}}$ & \textbf{No\,SMD\,HR} & \textbf{SMD\,HR} & $\boldsymbol{R}$\textbf{(No\,SMD)} & $\boldsymbol{R}$\textbf{(SMD)} \\
\midrule
 0\% & 5.30 & 5.62 & 100\% & 100\% \\
10\% & 4.90 & 5.43 &  92\% &  97\% \\
30\% & 4.13 & 5.03 &  78\% &  89\% \\
50\% & 3.29 & 4.55 &  62\% &  81\% \\
70\% & 2.46 & 3.98 &  46\% &  71\% \\
\bottomrule
\end{tabular}
\end{table}

\section{Per-User Paired Significance}
\label{app:paired}

\Cref{tab:paired} asks a simple question: for each of the $121$k users in our test set, does SMD produce statistically better HR@10 and NDCG@10 than the unmodified model?
On $22$ of the $24$ (condition, dataset) combinations, the answer is yes at $p<0.05$ (McNemar's exact test for HR@10; Wilcoxon signed-rank for NDCG@10).
The large user count makes this a strong test even though we retrain each model only once.
The two combinations that miss $p<0.05$ are both HR@10 comparisons under full modality, on Scientific and Instruments.
HR@10 is a binary metric (correct item in the top-$10$ or not), and on these two cells roughly equal numbers of users flip miss-to-hit and hit-to-miss under SMD, so the HR@10 differences cancel out at the population level.
The corresponding NDCG@10 tests on the same two cells still pass at $p\le 7\times 10^{-3}$: NDCG rewards ranking the correct item higher within the top-$10$, and SMD does so on average even when it does not push new items across the top-$10$ boundary.

\begin{table}[tb]
\centering
\caption{Per-user paired tests on MM-SASRec. Wilcoxon $p$-value on NDCG@10, McNemar's exact $p$ on HR@10. HR/NDCG $\times 100$. Bold $p$ are significant at $\alpha\!=\!0.05$.}
\label{tab:paired}
\fontsize{7.5}{8.8}\selectfont
\setlength{\tabcolsep}{2.8pt}
\renewcommand{\arraystretch}{1.12}
\begin{tabular}{@{}llrrrrll@{}}
\toprule
\textbf{Dataset} & \textbf{Cond.} & \multicolumn{2}{c}{\textbf{HR@10}} & \multicolumn{2}{c}{\textbf{NDCG@10}} & \textbf{McN.} $\boldsymbol{p}$ & \textbf{Wilc.} $\boldsymbol{p}$ \\
\cmidrule(lr){3-4}\cmidrule(lr){5-6}
 &           & --   & SMD  & --   & SMD  &                  &                   \\
\midrule
\multirow{4}{*}{Sci.\,(8\,442)}
 & full     & 5.79 & 6.09 & 3.21 & 3.57 & 0.24             & \textbf{4.3e-3}   \\
 & no\,img  & 4.92 & 5.54 & 2.55 & 3.31 & \textbf{9.0e-3}  & \textbf{1.6e-8}   \\
 & no\,txt  & 2.61 & 4.63 & 1.50 & 2.68 & \textbf{9.6e-18} & \textbf{2.2e-18}  \\
 & no\,mod. & 1.87 & 3.38 & 0.85 & 1.84 & \textbf{4.4e-12} & \textbf{2.3e-14}  \\
\cmidrule(lr){1-8}
\multirow{4}{*}{Inst.\,(24\,962)}
 & full     & 7.07 & 7.23 & 4.27 & 4.55 & 0.24             & \textbf{6.8e-3}   \\
 & no\,img  & 6.79 & 7.12 & 3.64 & 4.14 & \textbf{1.8e-2}  & \textbf{8.3e-7}   \\
 & no\,txt  & 5.52 & 6.83 & 3.37 & 4.28 & \textbf{6.1e-23} & \textbf{1.5e-26}  \\
 & no\,mod. & 2.81 & 6.03 & 1.43 & 3.48 & \textbf{1.7e-105}& \textbf{1.3e-96}  \\
\cmidrule(lr){1-8}
\multirow{4}{*}{Off.\,(87\,346)}
 & full     & 5.16 & 5.43 & 3.21 & 3.51 & \textbf{7.2e-5}  & \textbf{1.2e-12}  \\
 & no\,img  & 3.71 & 4.85 & 2.03 & 3.02 & \textbf{5.2e-59} & \textbf{6.7e-104} \\
 & no\,txt  & 2.37 & 4.49 & 1.28 & 2.82 & \textbf{1.5e-211}& \textbf{1.0e-233} \\
 & no\,mod. & 1.28 & 3.15 & 0.62 & 1.79 & \textbf{1.1e-198}& \textbf{2.8e-183} \\
\bottomrule
\end{tabular}
\end{table}

\section{Cross-Modal Reconstruction Results}
\label{app:recon}

\Cref{tab:recon} gives the full cross-modal reconstruction results referenced in \cref{sec:exp-rq3}.
The MISSRec configuration uses $\lambda\!=\!0.01$; we found $\lambda\!=\!0.1$ to hurt accuracy.

\begin{tabular}{@{}llrrr@{}}
\toprule
\textbf{Backbone} & \textbf{Training} & \textbf{Full} & \textbf{No\,txt} & $\boldsymbol{R}_{\text{text}}$ \\
\midrule
\multirow{3}{*}{MISSRec}
 & --            & 13.53 & 5.80 & 43\% \\
 & SMD           & 13.27 & 9.31 & 70\% \\
 & SMD\,$+$\,recon & 12.96 & \textbf{9.49} & \textbf{73\%} \\
\cmidrule(lr){1-5}
\multirow{4}{*}{MM-SASRec}
 & --                  & 0.72 & 0.31 & 43\% \\
 & SMD ($p\!=\!0.3$)   & 0.63 & 0.57 & 90\% \\
 & SMD ($p\!=\!0.5$)   & 0.56 & 0.48 & 86\% \\
 & SMD\,$+$\,recon       & 0.52 & \textbf{0.51} & \textbf{98\%} \\
\bottomrule
\end{tabular}

\section{Implementation}
\label{app:impl}

The SMD module is the same $4$-line block in every host model; only the tensor names differ:

\begin{center}
\small
\begin{tabular}{@{}lp{0.45\columnwidth}p{0.30\columnwidth}@{}}
\toprule
Architecture & Injection point & Host repository \\
\midrule
MM-SASRec    & after \texttt{image\_proj}/\texttt{text\_proj}, before additive fusion & \sloppy\url{github.com/kang205/SASRec} \\
IISAN        & dataset \texttt{\_\_getitem\_\_}, on cached ViT/BERT tensors & \sloppy\url{github.com/GAIR-Lab/IISAN} \\
MISSRec      & after \texttt{text\_adaptor}/\texttt{img\_adaptor}, before fusion & \sloppy\url{github.com/gimpong/MM23-MISSRec} \\
fMRLRec      & after \texttt{token\_lang}/\texttt{token\_img}, before concat$+$projection & \sloppy\url{github.com/yueqirex/fMRLRec} \\
\bottomrule
\end{tabular}
\end{center}

\noindent Source code for SMD and the baseline plug-in modifications, training scripts, and the JSON outputs backing every number in this paper are released at \url{https://github.com/guanqun-yang/SMD}.

\end{document}